\documentclass{PoS}
\usepackage{epsfig}
\usepackage{bm}
\usepackage{amssymb}
\usepackage{fontenc}
\usepackage{times}
\usepackage{mathptmx}
\usepackage{graphicx}

\title{New results with colour-sextet quarks}

\ShortTitle{New results with colour-sextet quarks}

\author{\speaker{D.~K.~Sinclair}%
\thanks{This research was supported in part by US Department of Energy
contract DE-AC02-06CH11357, and in part under a Joint Theory Institute
(JTI) grant.}\\
        HEP Division, Argonne National Laboratory, 9700 South Cass Ave.,
        Argonne, IL, 60439, USA 
        \\
        E-mail: \email{dks@hep.anl.gov}}

\author{J.~B.~Kogut
\thanks{Supported in part by a National Science Foundation grant 
NSF PHY03-04252.}\\
Department of Energy, Division of High Energy Physics, Washington, DC 20585,
USA\\
and\\
Dept. of Physics -- TQHN, Univ. of Maryland, 82 Regents Dr., College
Park, MD 20742, USA\\
        E-mail: \email{jbkogut@umd.edu}}

\abstract{We study QCD with 2 and 3 flavours of colour-sextet quarks. The
2-flavour theory is a candidate Walking Technicolor theory. Since we are
attempting to distinguish whether this theory is walking or conformal, we also
study the 3-flavour theory, which is believed to be conformal, for comparison.
We simulate lattice QCD with 2 and 3 flavours of colour-sextet staggered quarks
at finite temperatures to determine the scales of confinement and 
chiral-symmetry breaking from the positions of the deconfinement and
chiral-symmetry restoration transitions. Unlike the case with fundamental
quarks, these transitions are far apart. For 2 flavours the values of
$\beta=6/g^2$ for both transitions increase as $Ta$ is decreased from
$\frac{1}{4}$ to $\frac{1}{6}$ to $\frac{1}{8}$, as expected for a theory whose
coupling runs to smaller values as the lattice spacing is decreased. However,
for the chiral transition, the increase in $\beta$ between $Ta=\frac{1}{4}$ and
$Ta=\frac{1}{6}$ is much larger than the increase between $Ta=\frac{1}{6}$ and
$Ta=\frac{1}{8}$. This suggests that between $Ta=\frac{1}{4}$ and 
$Ta=\frac{1}{6}$ we are at strong coupling where the theory is effectively 
quenched, while between $Ta=\frac{1}{6}$ and $Ta=\frac{1}{8}$ we are emerging
into the weak coupling regime. It will require even smaller $Ta$ values to
determine whether the running of the chiral-transition coupling is controlled by
asymptotic freedom and the theory walks, or if it reaches a non-zero limit when
the transition becomes a bulk transition and the theory is conformal. The
3 flavour case at $Ta=\frac{1}{4}$ and $Ta=\frac{1}{6}$ behaves similarly to
the 2 flavour case. Since this theory is expected to be conformal, the
interpretation that we are seeing strong-coupling behaviour, inaccessible from
the weak-coupling limit (continuum) is the most likely interpretation.
}

\FullConference{The XXVIII International Symposium on Lattice Field Theory, 
                Lattice2010\\
		June 14-19, 2010\\
		Villasimius, Italy}

\begin{document}

\section{Introduction}

Technicolor theories are QCD-like theories where the techni-pions play the
r\^{o}le of the Higgs field, giving masses to the $W$ and $Z$
\cite{Weinberg:1979bn,Susskind:1978ms}. If the choice of
gauge fields and fermions is such that the coupling constant evolves very 
slowly -- walks -- phenomenological problems can be avoided when this 
Technicolor theory is extended to also give masses to the quarks and leptons
\cite{Holdom:1981rm,Yamawaki:1985zg,Akiba:1985rr,Appelquist:1986an}. 

QCD with $1\frac{28}{125} \leq N_f < 3\frac{3}{10}$ flavours of massless 
colour-sextet quarks is expected to be either a walking or conformal field 
theory. (For a good summary of what is known from perturbation theory about 
$SU(N)$ gauge theories with $N_f$ fermions in various representations, see
\cite{Dietrich:2006cm}.)
For $N_f=3$, conformal behaviour is expected. The $N_f=2$ theory could,
a priori, exhibit either behaviour. We simulate lattice QCD with $N_f=2$ 
colour-sextet quarks to determine which of these two options it chooses. We
also simulate the $N_f=3$ theory for comparison. 

Our earlier studies of the $N_f=2$ theory on lattices with $N_t=4,6$ indicated
that the well-separated deconfinement and chiral-symmetry restoration 
transitions move to weaker couplings as $N_t$ increases from $4$ to $6$ as
expected of the finite temperature transitions of an asymptotically free 
theory \cite{Kogut:2010cz}. 
This seemed to favour the `walking' scenario. We are now extending
these studies to $N_t=8$. Again both transitions move to weaker couplings 
than those for $N_t=6$. However, for the chiral transition, the decrease from
$N_t=6$ to $N_t=8$ is far less than that from $N_t=4$ to $N_t=6$. This raises
the possibility that the coupling will tend to a finite limit as 
$N_t \rightarrow \infty$. If so, the transition is a bulk transition, and the
theory is conformal. Otherwise, if this critical coupling vanishes in this
limit, the transition remains a finite temperature transition and the theory
walks. It will be necessary to simulate even larger $N_t$ to distinguish 
between these two possibilities.

We are also performing finite temperature simulations of the $N_f=3$ theory 
on $N_t=4,6$ lattices. The behaviour of this theory appears very similar to 
that of the $N_f=2$ theory; the deconfinement and chiral-symmetry restoration 
transitions are widely separated, and move to significantly weaker couplings 
as $N_t$ increases from $4$ to $6$. The main difference is that all transitions
occur at stronger couplings than for $N_f=2$, as expected. Since this theory
is expected to be conformal, this suggests that the $N_t=4,6$ chiral
transitions are in the strong-coupled domain, which is separated from the
weakly-coupled conformal regime by a bulk transition, and is hence
inaccessible.

\section{$N_f=2$ simulations and results}

We are simulating lattice QCD with the Wilson plaquette action and 2 flavours
of (unimproved) staggered colour-sextet quarks using the RHMC algorithm. We
study thermodynamics by simulating on a lattice with temporal extent 
$N_t=1/(Ta)$ where $T$ is the temperature and $a$ is the lattice spacing. The
spatial extent of the lattice $N_s \gg N_t$ (We are interested in the limit
$N_s \rightarrow \infty$).

Our earlier work used lattices with $N_t=4$ and $6$. We are now simulating on
lattices with $N_t=8$. The results we present here are on $16^3 \times 8$
lattices, with quark masses $m_q=0.005$, $m_q=0.01$ and $m_q=0.02$. We perform
simulations with $\beta=6/g^2$ in the range $5.5 \le \beta \le 7.4$. Away from
the deconfinement and negative-to-complex Wilson-line transitions, we perform
runs of 10,000 length-1 trajectories for each $(\beta,m)$. Close to these
transitions we increase this to 50,000 trajectories. To check for finite size
effects we are also performing simulations with $m_q=0.005$ at large $\beta$s
on $24^3 \times 8$ lattices.

Just above the deconfinement transition, the Wilson line shows a strong 3-state
signal, where it preferentially orients itself in a direction close to that
of one of the cube roots of unity. We therefore bin our `data' into 3 bins
according to whether the argument of the Wilson line is closest to $0$,
$2\pi/3$ or $-2\pi/3$, up until the complex Wilson-line states disorder to
a negative Wilson-line state.

\begin{figure}[htb]
\parbox{2.9in}{
\epsfxsize=2.9in
\epsffile{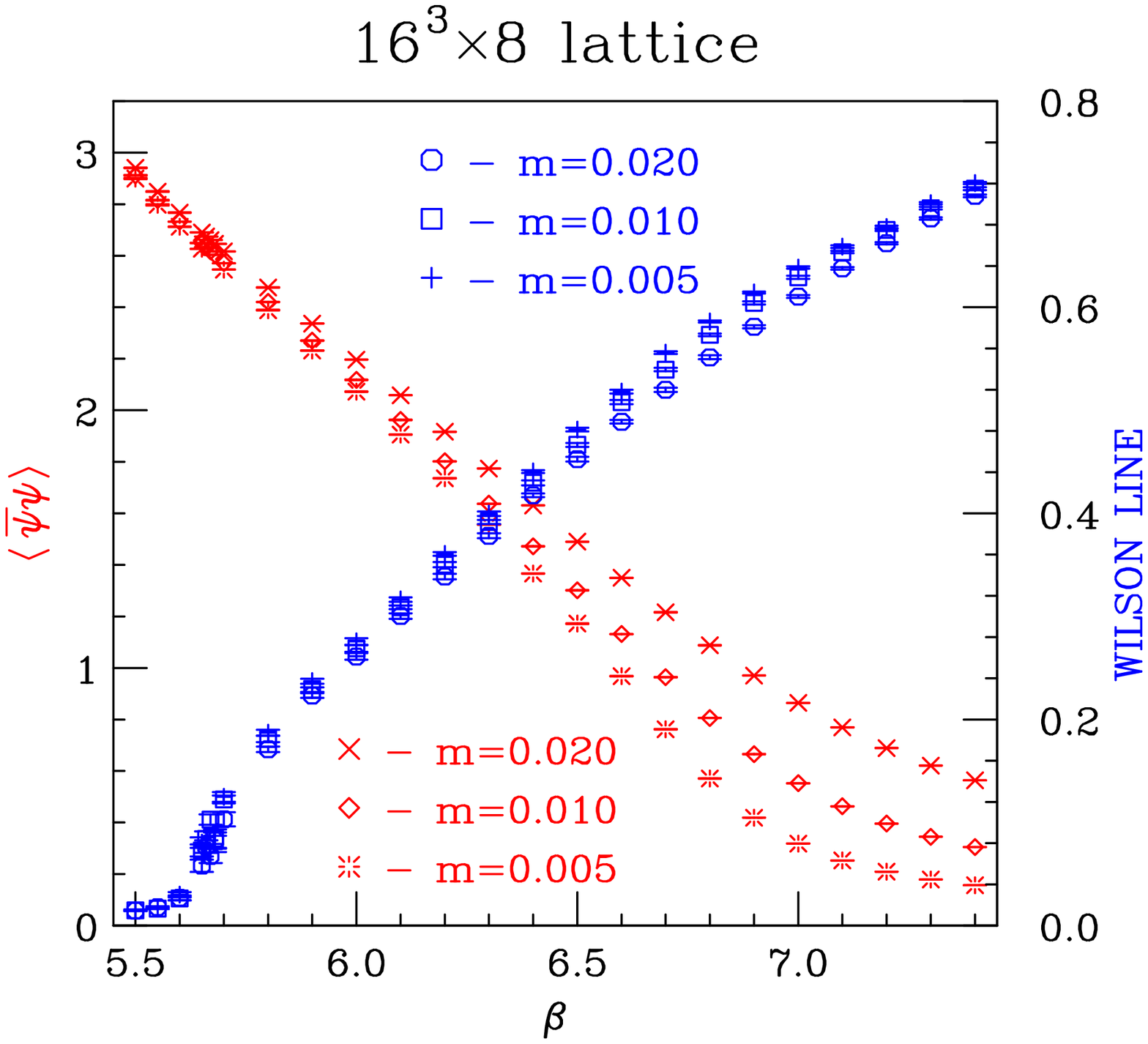}
\caption{Wilson Lines and chiral condensates for the positive Wilson Line
states, on a $16^3 \times 8$ lattice.}
\label{fig:rwil-psi_16x8}
}
\parbox{0.2in}{}
\parbox{2.9in}{
\epsfxsize=2.8in
\epsffile{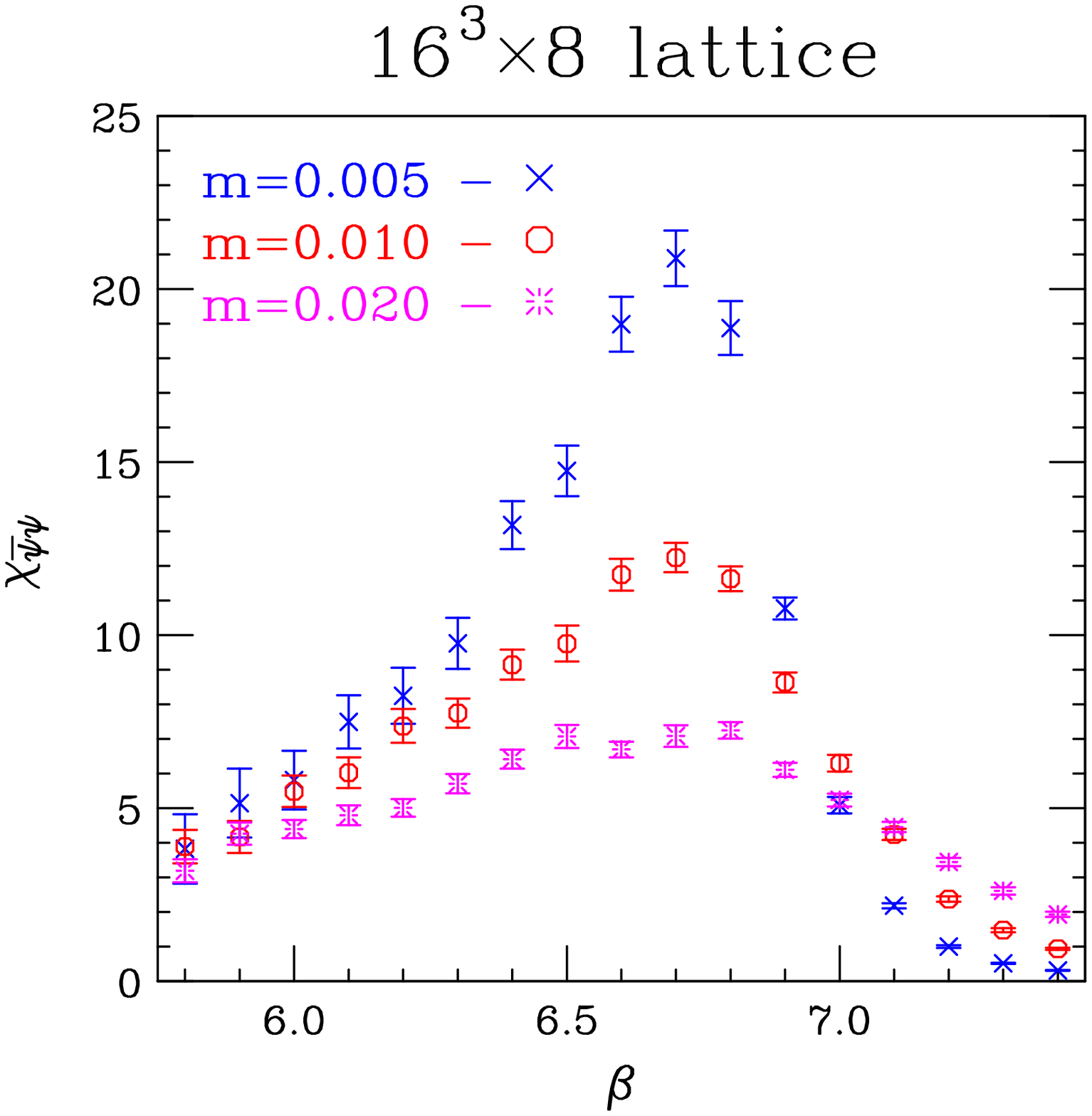}
\caption{Chiral susceptibilities on a $16^3 \times 8$ lattice.}
\label{fig:chi_16x8}
}
\end{figure}

Figure~\ref{fig:rwil-psi_16x8} shows the Wilson Lines and chiral condensates
($\langle\bar{\psi}\psi\rangle$) as functions of $\beta$ for each of the 3
quark masses, for our $N_f=2$ runs on a $16^3 \times 8$ lattice. This `data'
is for the real positive Wilson-Line bin only. The deconfinement transition is
identified as the point at which the magnitude of the Wilson Line jumps from
near zero to an appreciably larger value. From histograms of the magnitudes of
the Wilson lines, we estimate that the deconfinement $\beta$,
$\beta_d=5.665(10)$ for $m=0.02$ and $\beta_d=5.660(10)$ for $m=0.01$. 

In figure~\ref{fig:chi_16x8} we plot the disconnected chiral susceptibilities
for each of the 3 quark masses. For both $m=0.01$ and $m=0.005$, the peaks are
at $\beta \approx 6.7$, from which we estimate that in the chiral limit the
chiral phase transition is at $\beta=\beta_\chi=6.7(1)$. The states with
complex Wilson lines disorder into states with a real negative Wilson line
close to this chiral transition.

\section{$N_f=3$ simulations and results}

\subsection{$N_t=4$}

\begin{figure}[htb]
\parbox{2.9in}{
\epsfxsize=2.9in
\epsffile{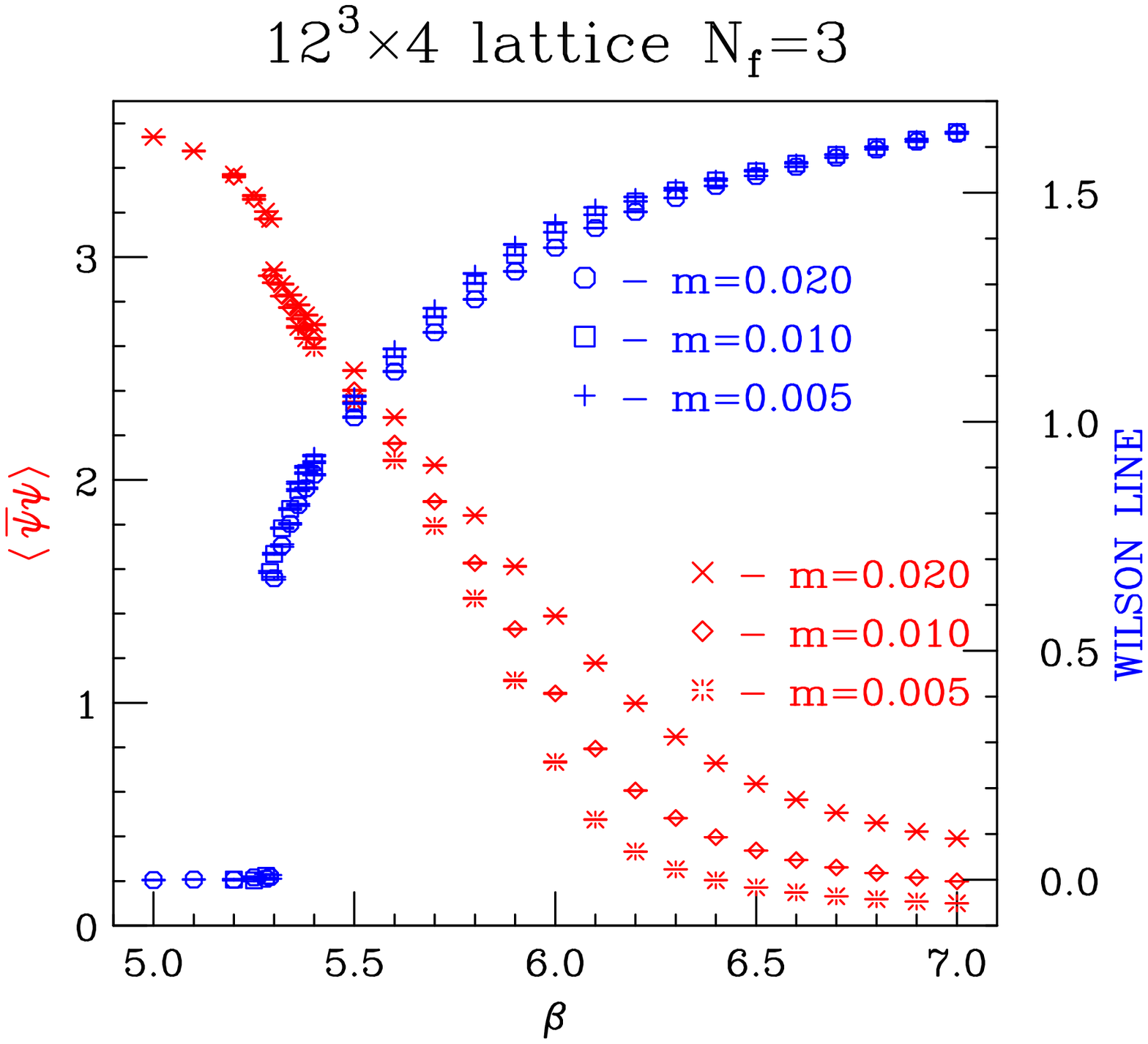}
\caption{Wilson Lines and chiral condensates for $N_f=3$ on a $12^3 \times 4$ 
lattice.}
\label{fig:wil-psi_12x4}
}
\parbox{0.2in}{}
\parbox{2.9in}{
\epsfxsize=2.9in
\epsffile{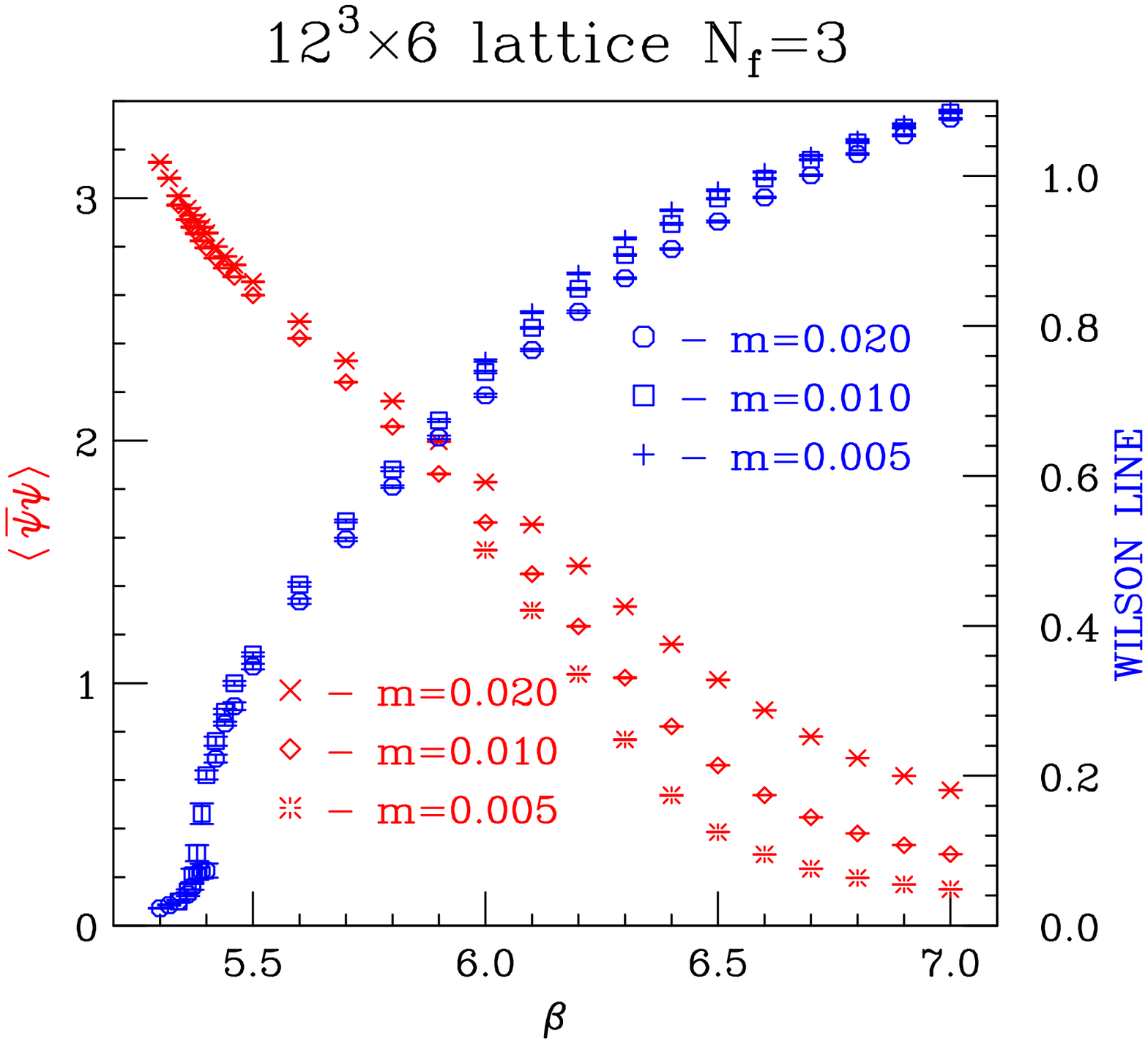}
\caption{Wilson Lines and chiral condensates for states with real positive
Wilson Lines for $N_f=3$ on a $12^3 \times 6$ lattice.}
\label{fig:wil-psi_12x6}
}
\end{figure}

We simulate lattice QCD with 3 flavours of colour-sextet quarks on 
$12^3 \times 4$ lattices, with quark masses $m=0.005$, $m=0.01$ and $m=0.02$.
$\beta$s are chosen covering the range $5 \le \beta \le 7$.
Figure~\ref{fig:wil-psi_12x4} shows the Wilson Lines and chiral condensates
for a series of runs starting from an ordered start (gauge fields $U=1$) at
$\beta=7$. Note that the deconfinement and chiral-symmetry restoration
transitions appear far apart.

From the time evolution of the Wilson lines close to the deconfinement 
transition, our preliminary estimates for the $\beta$ values for this transition
are $\beta_d(m=0.02)=5.295(5)$ and $\beta_d(m=0.01)=5.285(5)$. 
($\beta_d(m=0.005)$ has yet to be determined.) We estimate that the 
chiral-symmetry restoration phase transition at $m=0$ is at 
$\beta=\beta_\chi=6.0(1)$, from the peaks in the chiral susceptibilities. 
We are also performing a series of runs starting from a state where all gauge
fields $U=1$ except for the timelike fields on one timeslice which are set
equal to ${\rm diag}(1,-1,-1)$ at $\beta=7$. This forces the system into a
state with a negative Wilson line at large $\beta$s. From these runs we
observe that states having Wilson lines with arguments $\pm 2\pi/3$ disorder
into a state with a negative Wilson line above some $\beta$ in the range 
$5.5 < \beta²< 5.6$.

\subsection{$N_t=6$}

We simulate lattice QCD with 3 flavours of colour-sextet quarks on 
$12^3 \times—6$ lattices, with quark masses $m=0.005$, $m=0.01$ and $m=0.02$.
$\beta$s are chosen covering the range $5.3 \le‰\beta \le 7.0$. Above the
deconfinement transition the Wilson line exhibits a clear 3-state signal, so
we bin the `data' appropriately. Figure~\ref{fig:wil-psi_12x6} shows the
Wilson Lines and chiral condensates for the states with real positive Wilson
Lines for $N_f=3$ on a $12^3 \times 6$ lattice, for all 3 masses. Again, the
deconfinement and chiral-symmetry restoration transitions are well separated.

By looking at histograms of the magnitude of Wilson Lines we estimate that the
deconfinement transition for $m=0.02$ is at $\beta_d=5.41(1)$. For $m=0.01$ we
estimate that $\beta_d=5.395(5)$. We have yet to complete our simulations at
$m=0.005$ in the neighbourhood of the deconfinement transition. From the peaks
of the chiral susceptibilities for each of the 3 masses, we estimate that the
position of the $m=0$ chiral phase transition is at $\beta=\beta_\chi=6.3(1)$.
There also exists a state with a negative Wilson Line at large $\beta$. For
some $\beta$ in the range $6.1 \le \beta \le 6.2$ this undergoes a transition
to a state with its Wilson Line oriented in the direction of one of the complex
cube roots of unity. At $\beta$ values just above the deconfinement transition
we observe tunnelings between states with Wilson lines oriented in the 
directions of the 3 cube roots of unity.

\section{Discussion and Conclusions}

We are simulating the thermodynamics of lattice QCD with 2 flavours of
staggered colour-sextet quarks, whose chiral limit is a model of 
walking/conformal Technicolor. In addition we are simulating the 3-flavour 
theory, which is believed to have an infrared-stable fixed point, making it
a conformal field theory at zero quark mass.

We are extending our earlier $N_f=2$ simulations at $N_t=4,6$, to $N_t=8$. We
again find widely separated deconfinement and chiral-symmetry restoration
transitions. The $\beta$ values for these transitions are given in 
table~\ref{tab:N_feq2}.
\begin{table}[h]
\parbox{0.25in}{$\:$}
\parbox{2.5in}{
\centerline{
\begin{tabular}{|c|c|c|}
\hline
$N_t$          & $\beta_d$       & $\beta_\chi$             \\
\hline
4              &$\;$5.40(1)$\;$  &$\;$6.3(1)$\;$            \\
6              &$\;$5.54(1)$\;$  &$\;$6.6(1)$\;$            \\
8              &$\;$5.66(1)$\;$  &$\;$6.7(1)$\;$            \\
\hline
\end{tabular}
}
\caption{$N_f=2$ deconfinement and chiral transitions.}
\label{tab:N_feq2}
}
\parbox{0.5in}{$\:$}
\parbox{2.5in}{
\centerline{
\begin{tabular}{|c|c|c|}
\hline
$N_t$          & $\beta_d$        & $\beta_\chi$             \\
\hline                                             
4              &$\;$5.28(1)$\;$   &$\;$6.0(1)$\;$            \\
6              &$\;$5.39(1)$\;$   &$\;$6.3(1)$\;$            \\
\hline
\end{tabular}
}                   
\caption{$N_f=3$ deconfinement and chiral transitions.}
\label{tab:N_feq3}   
}
\end{table}

Both $\beta_d$ and $\beta_\chi$ increase significantly with increasing $N_t$.
However, the fact that the increase in $\beta_\chi$ between $N_t=6$ and $8$
is much smaller that that between $N_t=4$ and $6$ indicates that this is not
simply due to the expected running of the bare (lattice) coupling with lattice
spacing, controlled by asymptotic freedom. The most likely explanation is that
for $N_t=4$ and $6$, $\beta=\beta_\chi$ is in the strong-coupling domain. Here,
once the condensate forms, the quarks play little part in the running of the 
coupling constant, which now runs as in the quenched theory. Between $N_t=6$
and $8$ the theory is starting to emerge from the strong-coupling domain, and
the effects of the fermions are starting to be felt. We will therefore need to
simulate at even larger $N_t$ to see if the running of the coupling is now
controlled by the asymptotically free $g=0$ fixed point or if a bulk chiral
transition occurs before one reaches this fixed point. In the first case the
continuum theory is confining and chirally broken, but with a slowly running
-- walking -- coupling. In the second case the continuum theory is conformal.

DeGrand, Shamir and Svetitsky have studied QCD with 2 sextet Wilson quarks
\cite{Shamir:2008pb},
They also find it difficult to distinguish between walking and conformal
behaviour \cite{DeGrand:2010na}. 
At finite temperature, they do not, however, see any separation 
between the deconfinement and chiral transitions \cite{DeGrand:2008kx}. 
The Lattice Higgs Collaboration%
\footnote{Z.~Fodor, K.~Holland, J.~Kuti, D.~Nogradi, C.~Schroeder.}
have been studying QCD with 2 sextet improved staggered quarks
at zero temperatures. Their preliminary results are consistent with the
walking scenario \cite{Kuti}.

We are also simulating the thermodynamics of lattice QCD with 3 flavours of 
staggered colour-sextet quarks. Since this theory is expected to be conformal,
it is interesting to compare its behaviour with that of the 2-flavour model.
In our present simulations with $N_t=4,6$, the results look very similar to
the $N_f=2$ theory except that the transitions are (as expected) moved to
smaller $\beta$s. We have tabulated these results in table~\ref{tab:N_feq3}.
The deconfinement and chiral transitions are well separated and move to
significantly higher $\beta$ values as $N_t$ is increased from $4$ to $6$.
Since this theory is expected to be conformal, the large increase in 
$\beta_\chi$ is almost certainly quenched scaling in the strong-coupling
regime. In fact, $\beta_\chi$ at $N_t=4$ is close to $\beta_d$ for the quenched 
theory at $N_t=8$ while that at $N_t=6$ is close to the quenched $\beta_d$ at
$N_t=12$ \cite{Gottlieb:1985ug,Christ:1985wx,Boyd:1996bx}. 
Hence $\beta_\chi$ does appear to evolve with quenched dynamics.
Here we would expect that, once we move to large enough $N_t$, $\beta_\chi$
will approach a fixed finite value representing a bulk transition separating
the strongly coupled chirally-broken phase from the weak-coupling conformal
phase.

Above the deconfinement transition we see a 3-state signal, the remnant of the
now-broken $Z_3$ centre symmetry, for both $N_f=2$ and $N_f=3$. At an even
higher $\beta$, the 2 states with complex Wilson Lines disorder into a state
with a negative Wilson Line. The existence of states with Wilson Lines having
phases $\pm 2\pi/3$ and $\pi$ in addition to those with phase $0$
is predicted by Machtey and Svetitsky and observed in their simulations with 
Wilson quarks \cite{Machtey:2009wu}.

As well as needing lattices with larger $N_t$, we need to simulate at larger
spatial volumes, especially for $\beta$ in the neighbourhood of $\beta_\chi$
and at low quark masses, to understand finite size effects. We are currently
extending our $N_f=2$, $N_t=8$ simulations with $m=0.005$ to $24^3 \times 8$
lattices. $N_f=2$, $N_t=12$ simulations are planned, as are $N_f=3$, $N_t=8$
simulations. These will concentrate on the vicinity of the chiral transition.

Smaller quark masses are needed to be certain that we can access
the chiral limit. Zero temperature simulations are needed to study hadron 
spectra, $f_\pi$, interquark potential(s)..., to better determine whether these
theories are walking or conformal. Such simulations will also allow us to study
the flavour-symmetry (`taste') breaking from the use of staggered quarks, at
relevant $\beta$ values. Direct measurement of the running of the coupling
constant should be performed, along with measurement of the anomalous 
dimension of the chiral condensate.

\section*{Acknowledgements}

The simulations reported here were performed on the Fusion Cluster at the LCRC
at Argonne, the Franklin Cray XT4, the Hopper Cray XT5, and the Carver/Magellan
Cluster at NERSC, and the Kraken Cray XT5 at NICS.


\begin{thebibliography}{99}


\bibitem{Weinberg:1979bn}
  S.~Weinberg,
  Phys.\ Rev.\  D {\bf 19}, 1277 (1979).

\bibitem{Susskind:1978ms}
  L.~Susskind,
  Phys.\ Rev.\  D {\bf 20}, 2619 (1979).


\bibitem{Holdom:1981rm}
  B.~Holdom,
  Phys.\ Rev.\  D {\bf 24}, 1441 (1981).

\bibitem{Yamawaki:1985zg}
  K.~Yamawaki, M.~Bando and K.~i.~Matumoto,
  Phys.\ Rev.\ Lett.\  {\bf 56}, 1335 (1986).

\bibitem{Akiba:1985rr}
  T.~Akiba and T.~Yanagida,
  Phys.\ Lett.\  B {\bf 169}, 432 (1986).

\bibitem{Appelquist:1986an}
  T.~W.~Appelquist, D.~Karabali and L.~C.~R.~Wijewardhana,
  Phys.\ Rev.\ Lett.\  {\bf 57}, 957 (1986).


\bibitem{Dietrich:2006cm}
  D.~D.~Dietrich and F.~Sannino,
  Phys.\ Rev.\  D {\bf 75}, 085018 (2007)
  [arXiv:hep-ph/0611341].


\bibitem{Kogut:2010cz}
  J.~B.~Kogut and D.~K.~Sinclair,
  Phys.\ Rev.\  D {\bf 81}, 114507 (2010)
  [arXiv:1002.2988 [hep-lat]].


\bibitem{Shamir:2008pb}
  Y.~Shamir, B.~Svetitsky and T.~DeGrand,
  Phys.\ Rev.\  D {\bf 78}, 031502 (2008)
  [arXiv:0803.1707 [hep-lat]].

\bibitem{DeGrand:2010na}
  T.~DeGrand, Y.~Shamir and B.~Svetitsky,
  arXiv:1006.0707 [hep-lat].

\bibitem{DeGrand:2008kx}
  T.~DeGrand, Y.~Shamir and B.~Svetitsky,
  Phys.\ Rev.\  D {\bf 79}, 034501 (2009)
  [arXiv:0812.1427 [hep-lat]].


\bibitem{Kuti}
J.~Kuti, talk presented at Lattice 2010, Villasimius, Sardinia, Italy (2010).


\bibitem{Gottlieb:1985ug}
  S.~A.~Gottlieb, J.~Kuti, D.~Toussaint, A.~D.~Kennedy, S.~Meyer, 
  B.~J.~Pendleton and R.~L.~Sugar,
  Phys.\ Rev.\ Lett.\  {\bf 55}, 1958 (1985).

\bibitem{Christ:1985wx}
  N.~H.~Christ and A.~E.~Terrano,
  Phys.\ Rev.\ Lett.\  {\bf 56}, 111 (1986).

\bibitem{Boyd:1996bx}
  G.~Boyd, J.~Engels, F.~Karsch, E.~Laermann, C.~Legeland, M.~Lutgemeier and 
  B.~Petersson,
  Nucl.\ Phys.\  B {\bf 469}, 419 (1996)
  [arXiv:hep-lat/9602007].
 

\bibitem{Machtey:2009wu}
  O.~Machtey and B.~Svetitsky,
  Phys.\ Rev.\  D {\bf 81}, 014501 (2010)
  [arXiv:0911.0886 [hep-lat]].

\end{thebibliography}
\end{document}